# Asymptotic form of molecular continuum wave function for system of non-overlapping atomic potentials

A. S. Baltenkov

Arifov Institute of Electronics, 100125, Tashkent, Uzbekistan

**Abstract**
The asymptotic behavior of the molecular continuum wave function has been analyzed within a model of non-overlapping atomic potentials. It is been shown that the representation of the wave function far from a molecule as a plane wave and single spherical wave emitted by the molecular center cannot be corrected. Because of the multicenter character of the problem, the asymptotic form of the wave function must contain $N$ spherical waves with centers at the nuclei of the $N$ atoms that form the molecule. A method of partial waves for a spherically non-symmetrical target is considered for the simplest multicenter target formed by two non-overlapping potentials. The results are compared with those obtained within the single spherical wave approximation. It has been shown that the use of this approximation results in significant mistakes in differential and total cross sections of electron elastic scattering by a target.

1. Introduction

The multiple scattering (MS) methodology is one of the most popular theoretical constructions used for calculation of molecular continuum wave functions. The general ideas of this method were developed by Dill and Dehmer in paper [1] where "the multiple scattering technique for treating nonseparable eigenvalue problems with electron-scattering theory to construct continuum wave functions" was combined. The MS method [1] and its later modifications [2-9] are widely used now to calculate the cross sections of electron elastic scattering by molecules and those of molecular photoionization. Originally the MS method was used in molecular physics to calculate bound state eigenvalues [10]. For the calculation of the bound state wave functions their normalization is evident. The situation with the continuum wave functions is quite different. A choice of their normalization, i.e. the asymptotic behavior of the wave functions, requires to be analyzed. This moment is of great importance for the accuracy of any method of molecular continuum calculations to be estimated, particularly when one deals with differential cross sections of fixed-in-space targets, because these cross sections are extremely sensitive to the asymptotic behavior of the wave function.

One of the general assumptions of the methods [1-9] is based on the fact that the asymptotic form of the wave function far from the molecule is a sum of a plane wave plus a single spherical wave (SSW) emitted by the molecular center and therefore the radial parts of the electron wave functions outside so-called "molecular sphere" can be represented as a linear combination of the regular and irregular solutions of the Schrödinger equation for the potential that in this region "is taken to be spherical … about the molecular center" [1]. The coefficients of this linear combination are defined by the molecular phases of scattering. They are defined by the matching conditions of the continuum wave function on the surfaces of the atomic and molecular spheres. Hence, in the method [1] the solution of the problem of electron scattering by a spherically non-symmetrical potential is reduced without any grounds to the usual method of the partial waves for a spherical target. Proposed in paper [1], the recipes to build a continuum wave function outside the molecular sphere are considered as a matter-of-course and, as far as we know, they are beyond any doubt.

The general problem of multiple wave scattering by a system of scatterers was investigated long before appearance of the above mentioned papers. The review of these studies is given in Refs. [11-17]. A classical physical picture of wave scattering is based on the Huygens-Fresnel principle. According to this principle, the initial wave interacts with each target center that becomes a source of the secondary spherical scattered waves and beyond the target there is a system of the spherical waves diverging from each of the centers, rather than a single spherical wave as proposed in the method [1]. Schematically the difference in these two physical pictures of the wave scattering is given in Fig. 1*a*, *b*. It is known that the interference of the spherical waves emitted by the spatially separated sources creates a diffraction pattern whose properties depend periodically on the ratio of the internuclear distance to the electron wavelength. If we suppose, as it is done in [1], that far from the system of the scattering centers there is a single spherical wave (Fig. 1*b*) then the phenomenon of electron diffraction by molecules as the interference of a few spherical waves becomes impossible at all. Therefore it is difficult expect that the SSW assumption can be the basis for correct description of differential cross sections of photoionization or elastic scattering.

As for periodic modulations in the total cross section of molecular photoionization which are interpreted as a consequence of electron diffraction (see for example [18]), they have the completely different nature. Their appearance in the photoionization total cross section "is due to detailed properties of valence orbitals" [18], i.e., connected with the multicenter structure of the molecule initial state wave function rather than with interference of the secondary waves in the continuum. These undulations are qualitatively reproduced even in the simplest picture of the wave scattering Fig. 1*c*. In the Born approximation of zeroth order the wave function of the molecular continuum is described by a plane wave [18] that has no diverging spherical waves. Consequently, in this approximation the interference of secondary waves, i.e. electron diffraction in its classical understanding, is out of the question.

No clear understanding about what the picture of scattering by a system of atomic potentials is and what electron diffraction by a molecule is: interference of spherical waves in the continuum or periodic modulations in the total photoionization cross sections leads sometimes to mixing the pictures *a*, *b*, *c*. So, in Ref. [19] we read "…Cohen and Fano [18] discussed the role of interference in the photoelectron spectra of valence electrons (*using Fig. 1c*). Their theme was developed by Dehmer and Dill [1] into the K-shell spectroscopy of diatomic molecules (*Fig. 1b*). The idea behind it is sketched in *Fig. 1a*". (Here italics supplied).

In the connection with the above stated a question arises: Is it possible to adapt the method of partial waves for the case of a multicenter target keeping the Huygens-Fresnel picture of the scattering process according to which far from the target there is a system of the secondary waves diverging from each of the centers? The positive answer to this question was given by Demkov and Rudakov in paper [20] where it was shown that the S-matrix method can be also applied to non-spherical potentials. In the present paper for a simple example of scattering of a slow particle by two short-range non-overlapping potentials we will analyze the special features of the partial wave method for non-spherical targets and compare the results obtained with those following from the SSW asymptotic. As in Ref. [1], we will consider the targets with fixed distances between atoms, i.e. we neglect the interaction of the electron with molecular vibration and rotation. These fine effects can be introduced at the later stage.

Note that the asymptotic behavior of the continuum wave function for multicenter problem was studied in the recent paper [21] where the asymptotic forms of the scattered wave for a system of three charged particles were derived. The analysis of this classical problem of three bodies is essentially complicated by the Coulomb interaction between particles ant their comparable masses. This paper considers the simpler problem of light particle (electron) scattering by two fixed-in-space heavy scatterers (atoms) with which the particle interacts by short-range forces.

The paper plan is as follows. The general formulas the partial waves method for non-spherical targets is based on are presented in Sec. 2. The presentation of these formulas is quite appropriate here because in the literature devoted to molecular scattering Refs. [20, 22] are not cited at all. The problem of slow particle scattering by two short-range potentials is considered in Sec. 3 and 4. First this problem is solved by calculating the scattering amplitude in the closed form with subsequent application of the optical theorem and then by the S-matrix method [20, 22] with the general formulas of Sec. 2. The both methods are shown to lead to identical general formulas for the differential and total cross sections of scattering. In Sec. 5 the same problem is solved in the SSW approximation. First the connection between the scattering amplitude and molecular phases in Ref. [1] is established. Then these phases are calculated from the exact scattering amplitude. It is shown that there are no phase functions capable of providing the identity of the scattering amplitude calculated by the S-matrix method [20, 22] with the amplitude in method [1]. In Sec. 6 the applicability range of the SSW approximation is estimated for the example of slow electron scattering by $N_2$ molecule. The conclusions are presented in Sec. 7.

2. Method of partial waves for non-spherical targets

Let us briefly describe at first the main ideas of paper [20]. It is known that the wave function describing elastic scattering of a particle by a spherically symmetrical potential is defined by the expression [23]

$$\psi_{\mathbf{k}}^{+}(\mathbf{r}) = 4\pi \sum_{l=0}^{\infty} R_{klm}(r) Y_{lm}^{*}(\mathbf{k}) Y_{lm}(\mathbf{r}), \qquad (1)$$

where the radial part of the wave function has the asymptotic form

$$R_{klm}(r \to \infty) \approx e^{i(\delta_l + \frac{\pi l}{2})} \frac{1}{kr} \sin(kr - \frac{\pi l}{2} + \delta_l). \qquad (2)$$

A molecular potential as a cluster of non-overlapping spherical potentials centered at the atomic sites is a non-spherical potential. In the Schrödinger equation with this potential it is impossible to separate the angular variables and present the wave function at an arbitrary point of space in the form of expansion in spherical functions (1). However, asymptotically at great distances from the molecule the wave function can be written as expansion in a set of other orthonormal functions $Z_\lambda(\mathbf{k})$:

$$\psi_{\mathbf{k}}^{+}(\mathbf{r} \to \infty) \approx 4\pi \sum_{\lambda} R_{k\lambda}(r) Z_{\lambda}^{*}(\mathbf{k}) Z_{\lambda}(\mathbf{r}) \qquad (3)$$

with the radial part of the wave function

$$R_{k\lambda}(r \to \infty) \approx e^{i(\eta_\lambda + \frac{\pi}{2}\omega_\lambda)} \frac{1}{kr} \sin(kr - \frac{\pi}{2}\omega_\lambda + \eta_\lambda). \qquad (4)$$

Here the index $\lambda$ numerates different partial functions similar to the quantum numbers $l$ and $m$ for the central field; $\omega_\lambda$ is the quantum number (parity) that is equal to the orbital moment $l$ for the spherical symmetry case; $\eta_\lambda(k)$ are the molecular phases. The explicit form of functions $Z_\lambda(\mathbf{k})$, naturally, depends on a specific type of the target field, particularly on the number of atoms forming the target and on mutual disposition of the scattering centers in space, *etc*. The functions $Z_\lambda(\mathbf{k})$, like the spherical functions $Y_{lm}(\mathbf{k})$, create an orthonormal system [20] and for this reason:

$$\int Z_\lambda^{*}(\mathbf{k}) Z_\mu(\mathbf{k}) d\Omega_k = \delta_{\lambda\mu}. \qquad (5)$$

The scattering amplitude for a non-spherical target, according to [20], is given by the following expression

$$F(\mathbf{k},\mathbf{k}') = \frac{2\pi}{ik} \sum_\lambda (e^{2i\eta_\lambda} - 1) Z_\lambda^*(\mathbf{k}) Z_\lambda(\mathbf{k}'). \tag{6}$$

The total elastic scattering cross section, i.e. the cross section integrated over all directions of momentum of the scattered electron **k'**, is defined by the formula

$$\sigma(\mathbf{k}) = \frac{(4\pi)^2}{k^2} \sum_\lambda |Z_\lambda(\mathbf{k})|^2 \sin^2 \eta_\lambda. \tag{7}$$

Of course, this cross section depends on the mutual orientation of incident electron momentum **k** and molecule axes. The cross section averaged over all the directions of momentum of incident electron **k** is connected with the molecular phases $\eta_\lambda(k)$ by the following formula

$$\overline{\sigma}(k) = \frac{4\pi}{k^2} \sum_\lambda \sin^2 \eta_\lambda. \tag{8}$$

In the case of a spherical symmetrical target the formula (8) exactly coincides with the known formula for the total scattering cross section. Indeed, in the case of the central field the index $\lambda$ is replaced by the quantum numbers $l$ and $m$. But the phase of scattering by the central field is independent of the magnetic number and therefore for the given value of the orbital moment $l$ it is necessary to summarize over all $m$. This results in the factor $(2l+1)$ under the summation sign in formula (8).

The partial wave (4) and molecular phases $\eta_\lambda(k)$ are classified, according to [20], by their behavior for low electron energies, i.e. for $k \to 0$. In this limit the particle wavelength is great as compared with the target size and the function $Z_\lambda(\mathbf{k})$ tends to some spherical function $Y_{lm}(\mathbf{k})$. The corresponding phase is characterized in this limit by the following asymptotic behavior: $\eta_\lambda(k) \to k^{2l+1}$.

3. Scattering of a slow particle by two short-range potentials. Optical theorem

For the molecular system that is created by two short-range potentials each of which is a source of the scattered *s*-waves, the molecular phase shifts $\eta_\lambda(k)$ and the functions $Z_\lambda(\mathbf{k})$ can be calculated in the explicit form [22]. This simplest multicenter system is a good example illustrating the method of partial waves for non-spherical targets formed by non-overlapping atomic potentials. It is important that the problem of slow particle scattering by this system of centers can be solved analytically and so it can serve as a touchstone to analyze correctness of different calculation methods.

Consider the scattering of a slow electron by two identical non-overlapping atomic potentials with the centers at $\mathbf{r} = \pm \mathbf{R}/2$. Beyond the action of the short-range potentials the wave function for this system has the form (see for example [16] where this function was used to describe scattering of slow mesons by deuterons)

$$\psi_\mathbf{k}^+(\mathbf{r}) = e^{i\mathbf{k}\cdot\mathbf{r}} + D_1(\mathbf{k})\frac{e^{ik|\mathbf{r}+\mathbf{R}/2|}}{|\mathbf{r}+\mathbf{R}/2|} + D_2(\mathbf{k})\frac{e^{ik|\mathbf{r}-\mathbf{R}/2|}}{|\mathbf{r}-\mathbf{R}/2|}, \quad |\mathbf{r}\pm\mathbf{R}/2| \geq \rho. \tag{9}$$

Here $\rho$ is the short-range potential radius and the coefficients at the spherical waves have the form [24]

$$D_1(\mathbf{k}) = \frac{ad - bd^*}{a^2 - b^2}; D_2(\mathbf{k}) = \frac{ad^* - bd}{a^2 - b^2};$$

$$a = e^{ikR}/R; \quad d = -e^{i\mathbf{k}\cdot\mathbf{R}/2}; \quad b = ik - k\cot\delta_0 = ik + q. \tag{10}$$

The phase $\delta_0(k)$ is the *s*-wave phase for scattering by each of the atomic potentials forming the target. The wave function (9) is the general solution of the Schrödinger equation and describes multiple scattering of a particle by two identical potentials. This function has the form of a superposition of plane wave plus *two* spherical *s*-waves generated by each of the target atoms. The function (9) corresponds to the Huygens-Fresnel pattern of scattering according to which the wave scattering by a system of the *N* centers is accompanied by generation of *N* secondary waves.

The amplitude of the slow particle scattering by the target is obtained by considering the asymptotic behavior of the wave function (9)

$$F(\mathbf{k},\mathbf{k}',\mathbf{R}) = \frac{2}{a^2 - b^2}\{b\cos[(\mathbf{k}-\mathbf{k}')\cdot\frac{\mathbf{R}}{2}] - a\cos[(\mathbf{k}+\mathbf{k}')\cdot\frac{\mathbf{R}}{2}]\}. \tag{11}$$

The scattering cross section is obtained from the amplitude (11) with the help of the optical theorem [23]

$$\sigma(k,\mathbf{R}) = \frac{4\pi}{k}\operatorname{Im}F(\mathbf{k}=\mathbf{k}',\mathbf{R}) = \frac{8\pi}{k}\operatorname{Im}\left[\frac{b - a\cos(\mathbf{k}\cdot\mathbf{R})}{a^2 - b^2}\right]. \tag{12}$$

We introduce the vector **R** in the argument of the cross section (12) to underline that we deal with the fixed-in-space molecule. The total cross section averaged over all the directions of momentum of incident electron **k** has the form:

$$\bar{\sigma}(k) = \frac{1}{4\pi}\int\sigma(k,\mathbf{R})d\Omega_k = \frac{8\pi}{k}\operatorname{Im}\left[\frac{b - aj_0(kR)}{a^2 - b^2}\right]. \tag{13}$$

Here $j_0(x) = \sin x/x$ is the spherical Bessel function.

Using the explicit expressions (10) for the functions *a* and *b*, we obtain the following formula for the averaged cross section

$$\bar{\sigma}(k) = \frac{4\pi}{k^2}\left\{\left[1 + \left(\frac{qR + \cos kR}{kR + \sin kR}\right)^2\right]^{-1} + \left[1 + \left(\frac{qR - \cos kR}{kR - \sin kR}\right)^2\right]^{-1}\right\}. \tag{14}$$

Pay attention to the following moment. The total cross section (13) contains the term $\sin kR/kR$ characteristic of diffraction phenomena (see about it, for example, [18]). Its appearance in this case is connected with interference of two *s*-waves in the continuum wave function (9). Following the ideas of Ref. [18] let us analyze the reason for appearance of the similar term in the total cross section of molecule photoionization. The *g*-ground state of two-atomic molecule is described by the wave function

$$\psi_g(\mathbf{r}) \sim u(\mathbf{r} + \mathbf{R}/2) + u(\mathbf{r} - \mathbf{R}/2). \tag{15}$$

The dipole matrix element corresponding to photoionization of this state in the zeroth Born approximation [18] is defined by the integral

$$D_g \sim \int e^{-i\mathbf{k}\cdot\mathbf{r}}(\mathbf{e}\cdot\nabla)\psi_g d\mathbf{r} = (\mathbf{e}\cdot\mathbf{k})\int e^{-i\mathbf{k}\cdot\mathbf{r}}\psi_g d\mathbf{r} = 2\cos(\mathbf{k}\cdot\mathbf{R}/2)\int e^{-i\mathbf{k}\cdot\mathbf{r}}u(\mathbf{r})d\mathbf{r}. \quad (16)$$

The matrix element module square averaged over all possible directions of the molecular axis is

$$\int |D_g|^2 d\Omega_R = 2|J|^2 (1+\sin kR/kR). \quad (17)$$

Here $J$ is the Fourier-transform of the function $u(r)$, i.e., the integral in equation (16). In the similar way we obtain for the $u$-ground state

$$\int |D_u|^2 d\Omega_R = 2|J|^2 (1-\sin kR/kR). \quad (18)$$

Thus, the appearance of the oscillating function $\sin kR/kR$ in the total cross section of molecule photoionization results from the translation symmetry of the valence orbitals of two atomic molecules and is not connected with diffraction of the waves emitted by the separate sources of photoelectrons.

4. Scattering of a slow particle by two short-rang potentials. Method of partial waves

Now we are solving the scattering problem by the method of partial waves for non-spherical targets [20, 22]. The scattering amplitude (11) we rewrite in the form

$$F(\mathbf{k},\mathbf{k'},\mathbf{R}) = -\frac{2}{a+b}\cos(\mathbf{k}\cdot\mathbf{R}/2)\cos(\mathbf{k'}\cdot\mathbf{R}/2) + \frac{2}{a-b}\sin(\mathbf{k}\cdot\mathbf{R}/2)\sin(\mathbf{k'}\cdot\mathbf{R}/2). (19)$$

According to [20], the amplitude (11) should be considered as the sum of two partial amplitudes. The first of them is written as

$$\frac{4\pi}{2ik}(e^{2i\eta_0}-1)Z_0(\mathbf{k})Z_0^*(\mathbf{k'}) = -\frac{2}{a+b}\cos(\mathbf{k}\cdot\mathbf{R}/2)\cos(\mathbf{k'}\cdot\mathbf{R}/2). \quad (20)$$

The second is defined by the following expression

$$\frac{4\pi}{2ik}(e^{2i\eta_1}-1)Z_1(\mathbf{k})Z_1^*(\mathbf{k'}) = \frac{2}{a-b}\sin(\mathbf{k}\cdot\mathbf{R}/2)\sin(\mathbf{k'}\cdot\mathbf{R}/2). \quad (21)$$

The reasons for which we assigned the indexes at the functions $Z_\lambda(\mathbf{k})$ the values $\lambda = 0,1$ will become understandable further. From formulas (20) and (21) after elementary transformations we obtain *two* molecular phases of scattering (the proper phases in [20])

$$\cot\eta_0 = \frac{\text{Re}[(a+b)^*]}{\text{Im}[(a+b)^*]} = -\frac{qR+\cos kR}{kR+\sin kR}, \quad \cot\eta_1 = \frac{\text{Re}[(a-b)^*]}{\text{Im}[(a-b)^*]} = -\frac{qR-\cos kR}{kR-\sin kR}.(22)$$

Substituting the phase shifts (22) in formulas (20) and (21), we obtain the functions $Z_\lambda(\mathbf{k})$ in the explicit form ($Z_\lambda(\mathbf{k})$ are the characteristic scattering amplitudes in [20]). They are defined by the following expressions

$$Z_0(\mathbf{k}) = \frac{\cos(\mathbf{k} \cdot \mathbf{R}/2)}{\sqrt{2\pi S_+}}, \quad Z_1(\mathbf{k}) = \frac{\sin(\mathbf{k} \cdot \mathbf{R}/2)}{\sqrt{2\pi S_-}}. \tag{23}$$

Here $S_\pm = 1 \pm j_0(kR)$. It is easy to make sure that the functions (23) obey the conditions (5). It is evident that the functions (23) are defined by the geometrical target structure, i.e. by the direction of the molecular axis $\mathbf{R}$ in the arbitrary coordinate system in which the electron momentum vectors before and after scattering are $\mathbf{k}$ and $\mathbf{k'}$, respectively.

Study now the asymptotical behavior of the wave function (3) and (4). For this we write the exponent $\exp(i\mathbf{k}\cdot\mathbf{r})$ in the formula (9) as expansion in functions $Z_\lambda(\mathbf{k})$:

$$e^{i\mathbf{k}\cdot\mathbf{r}} = \sum_\lambda c_\lambda Z_\lambda(\mathbf{k}). \tag{24}$$

Multiplying the both parts of this equality by $Z_\mu^*(\mathbf{k})$ and integrating over all angles of the vector $\mathbf{k}$, we obtain the following expressions for the coefficients of the expansion (24):

$$c_0 = \sqrt{\frac{2\pi}{S_+}}[j_0(k|\mathbf{r}+\mathbf{R}/2|) + j_0(k|\mathbf{r}-\mathbf{R}/2|)],$$

$$c_1 = -i\sqrt{\frac{2\pi}{S_-}}[j_0(k|\mathbf{r}+\mathbf{R}/2|) - j_0(k|\mathbf{r}-\mathbf{R}/2|)]. \tag{25}$$

At large distances from the target the expansion coefficients in equations (25) have the form

$$c_0(r \to \infty) \approx 4\pi \frac{\sin kr}{kr} Z_0^*(\mathbf{k}) \text{ and } c_1(r \to \infty) \approx -i4\pi \frac{\cos kr}{kr} Z_1^*(\mathbf{k}). \tag{26}$$

Consider now the asymptotical behavior of the partial wave with the index $\lambda = 0$. Taking into account the formulas (9), (20), (24) and (26), we write the corresponding partial wave from formula (3) in the form

$$4\pi R_{k0}(r) Z_0(\mathbf{k}) Z_0^*(\mathbf{k}) = \frac{4\pi}{kr}[\sin kr + \frac{1}{2i}(e^{2i\eta_0}-1)e^{ikr}]Z_0(\mathbf{k}) Z_0^*(\mathbf{k}). \tag{27}$$

From equation (27) immediately we obtain

$$R_{k0}(r \to \infty) = e^{i\eta_0}\frac{1}{kr}\sin(kr+\eta_0). \tag{28}$$

Making the same operations for the case $\lambda = 1$, we obtain for the second partial wave the following expression

$$R_{k1}(r \to \infty) = e^{i(\eta_1+\frac{\pi}{2})}\frac{1}{kr}\sin(kr-\frac{\pi}{2}+\eta_1). \tag{29}$$

Comparing these expressions with the general formula (4) we come to the following conclusions. If the electron states are characterized by a projection of the angular momentum on the $\mathbf{R}$ axis and by parity of the wave function relative to the reflection in the plane perpendicular to **R** and going through the middle of the inter-atomic distance, then the first of the partial waves (28) corresponds to the state $\Sigma_g$ and the second one (29) to $\Sigma_u$. The molecular phases $\eta_\lambda(k)$ can be classified by considering their behavior at $k \to 0$ [20]. In this limit the electron wavelength is much greater than the target size and the picture of scattering have to approach to the spherical symmetry one. Consider this limit transition in the formulas (22); we obtain: $\eta_0(k \to 0) \sim k$ and $\eta_1(k \to 0) \sim k^3$. Thus, the molecular phases behave similar to the *s* and *p* phases in the spherically symmetrical potential. By this is explained the choice of their indexes. The transition to the limit $k \to 0$ in formulas (23) gives instead of the functions $Z_\lambda(\mathbf{k})$ the well-known spherical functions

$$Z_0(\mathbf{k})_{k \to 0} \to \frac{1}{\sqrt{4\pi}} \equiv Y_{00}(\mathbf{k}), \ Z_1(\mathbf{k})_{k \to 0} \to \sqrt{\frac{3}{4\pi}} \cos \vartheta \equiv Y_{10}(\mathbf{k}). \tag{30}$$

Here $\vartheta$ is the angle between the vector **k** and axis **R**.

Finally, substituting the molecular phases (22) in the formula (8), we obtain the cross section of elastic scattering by the target under consideration [20, 22]

$$\bar{\sigma}(k) = \frac{4\pi}{k^2}[\sin^2 \eta_0 + \sin^2 \eta_1] = \frac{4\pi}{k^2}[(1 + \cot^2 \eta_0)^{-1} + (1 + \cot^2 \eta_1)^{-1}]. \tag{31}$$

The same result was obtained above, equation (14), with the help of the optical theorem.

Summarize the results obtained with the method of partial waves for a target formed by non-overlapping atomic potentials. For such targets the molecular phases of scattering and the functions $Z_\lambda(\mathbf{k})$ can be found explicitly. The phases of molecular scattering, as the atomic ones, are the functions of electron momentum $k = |\mathbf{k}|$ only. The form of the functions $Z_\lambda(\mathbf{k})$ is defined by the structure of a target and its orientation in space. The number of non-zero molecular phases in this case is equal to two. This is connected with the fact that each of these two scattering centers is a source of *s*-spherical waves only, which is valid for the case of low electron energy. If the scattering by each of these centers would be accompanied by generation of spherical waves with non-zero orbital moments (this case was considered in recent paper [25]) then the number of non-zero molecular phases $\eta_\lambda(k)$ would be greater.

It is simple to calculate the functions $Z_\lambda(\mathbf{k})$ and the scattering phases $\eta_\lambda(k)$ when one knows the exact wave function (9). On the other hand, if this function is known, as it is the case for a system of non-overlapping potentials, the scattering amplitude can be obtained in the closed form (11) and the cross section can be found with the help of the optical theorem, and therefore there is necessity to resort to the method of partial waves. However, for non-spherical potentials different from muffin-tin-potentials, i.e. when the model of non-overlapping potential becomes inapplicable, the use of the partial wave method [20] makes it possible to separate in the explicit form the scattering dynamics contained in the molecular phases $\eta_\lambda(k)$ from the kinematics of the process and from target structure defined by the functions $Z_\lambda(\mathbf{k})$.

5. The continuum wave function with a single spherical wave

Solve now the same problem in accordance with the ideas of the method [1]. For this purpose we encircle a target with the molecular sphere as shown in Fig. 1*b*. Beyond this sphere (Region III) at great distances from the molecular center the scattering wave function is the sum of the incident plane wave and single outgoing spherical wave diverging from the molecular center (see equation (31) in Ref. [1])

$$\psi_{\mathbf{k}}^{+}(\mathbf{r})_{r\to\infty} \approx e^{i\mathbf{k}\cdot\mathbf{r}} + A(\mathbf{k}\cdot\mathbf{k}')\frac{e^{ikr}}{r} \qquad (32)$$

The partial *lm*-part of asymptotic of the wave function (32) has the form of a linear combination of regular and irregular solutions of the Schrödinger equation (see equations (8) and (20) of Ref. [1]) with the asymptotic behavior (2). The coefficients of this combination are defined by the scattering phases $\Delta_l$. Following [23], let us establish the connection between the scattering amplitude and the phase shifts $\Delta_l$. The amplitude $A(\mathbf{k}\cdot\mathbf{k}')$ is a function of the scalar product of the vectors $\mathbf{k}$ and $\mathbf{k}' = k\mathbf{r}/r$. Therefore, it can be written as an expansion in the Legendre polynomials

$$A(\mathbf{k}\cdot\mathbf{k}') = \sum_l (2l+1) f_l(k) P_l(\vartheta). \qquad (33)$$

The plane wave in function (32) can be written in the same way. Substituting these both expansions into (32), we obtain, taking into account equations (1) and (2), the following equation for the partial amplitude $f_l(k)$

$$e^{i\frac{\pi l}{2}}[e^{i\Delta_l}\sin(kr-\frac{\pi l}{2}+\Delta_l) - \sin(kr-\frac{\pi l}{2})] = k f_l(k) e^{ikr}. \qquad (34)$$

From here after the elementary transformations we obtain:

$$f_l(k) = \frac{1}{2ik}(e^{2i\Delta_l} - 1). \qquad (35)$$

Consequently, the total scattering amplitude $A(\mathbf{k}\cdot\mathbf{k}')$ in [1] coincides with the amplitude of scattering by a spherically symmetric potential and has the form [23]

$$A(\mathbf{k}\cdot\mathbf{k}') = \frac{2\pi}{ik}\sum_{l,m}(e^{2i\Delta_l}-1)Y_{lm}(\mathbf{k})Y_{lm}^*(\mathbf{k}'). \qquad (36)$$

This should be expected because one of the main assumptions in the method [1] is that the beyond the molecular sphere "the potential $V_{III}$ is taken to be spherical" [1]. The difference from the spherical case is that the scattering phases $\Delta_l$ are defined in [1] from the matching conditions for the wave function at the borders of atomic and molecular spheres.
    If the formulas (32) and (36) being correct for the spherically symmetrical potentials are valid for our problem then after the whole infinite set of the molecular phases is taken into account, the scattering amplitudes (11) and (36) have to coincide with each other, which follows from the essence of the partial expansion (36). The equality of these amplitudes is actually the equation for molecular phases:

$$F(\mathbf{k},\mathbf{k}',\mathbf{R}) = \frac{2\pi}{ik}\sum_{l,m}(e^{2i\Delta_l}-1)Y_{lm}(\mathbf{k})Y_{lm}^*(\mathbf{k}'). \qquad (37)$$

The amplitude $F(\mathbf{k},\mathbf{k}',\mathbf{R})$ is the function of three vectors and therefore it can be always presented as an expansion in tripolar spherical harmonics [26]. These harmonics are an irreducible tensor product of the three spherical functions $Y_{lm}(\mathbf{k})$, $Y_{lm}(\mathbf{k}')$ and $Y_{lm}(\mathbf{R})$.

Consequently, according to equation (37), the scattering phases $\Delta_l$ should be the functions not only of $k = |\mathbf{k}|$ but also of the spherical function $Y_{lm}(\mathbf{R})$. This is the principal difference between the partial wave method [20] in which the molecular phases are independent of target structure but are the functions of electron energy $\varepsilon$ only.

Integrating the both parts of equation (37) over the spherical angles of vectors $\mathbf{k}$ and $\mathbf{k'}$, we obtain the following expression

$$J_{lm} = \iint F(\mathbf{k},\mathbf{k'}) Y^*_{lm}(\mathbf{k}) Y_{lm}(\mathbf{k'}) d\Omega d\Omega' = \frac{2\pi}{ik}(e^{2i\Delta_l} - 1). \tag{38}$$

To calculate the integral $J_{lm}$ in the left side of equation (38), we write the cosines and sines in equation (11) as exponents and expand them as a series in spherical harmonics. Then integrating over spherical angles of vectors $\mathbf{k}$ and $\mathbf{k'}$, we obtain the following equation for phase shifts

$$J_{lm} = 32\pi^2 j_l^2(kR/2) |Y_{lm}(\mathbf{R})|^2 \frac{b - (-1)^l a}{a^2 - b^2} = \frac{2\pi}{ik}(e^{2i\Delta_l} - 1). \tag{39}$$

Thus, information on the exact scattering amplitude (11) allows unambiguous determination of the *infinite number of the phases* $\Delta_l$ ensuring the equality of the scattering amplitudes $A(\mathbf{k},\mathbf{k'})$ and $F(\mathbf{k},\mathbf{k'},\mathbf{R})$. According to equation (39), these phases should also depend on the magnetic quantum number *m*. Indirect indication to this feature of molecular phases for the SSW asymptotic we find, for example, in Ref. [27].

The phase shifts defining a linear combination of regular and irregular solutions of the wave equation in Region III, "form the real symmetric K matrix" [1]. Hence, the phase shifts $\Delta_l$ in equation (39) are the real numbers. Taking it into account and separating the real and imaginary parts in the both sides of equation (39), we obtain the following equations

$$\frac{4\pi}{k}\cos\Delta_l \sin\Delta_l = 32\pi^2 j_l^2(kR/2) |Y_{lm}(\mathbf{R})|^2 \operatorname{Re}\frac{b - (-1)^l a}{a^2 - b^2},$$

$$\frac{4\pi}{k}\sin^2\Delta_l = 32\pi^2 j_l^2(kR/2) |Y_{lm}(\mathbf{R})|^2 \operatorname{Im}\frac{b - (-1)^l a}{a^2 - b^2}. \tag{40}$$

From here we have the following formulas for $\cot\Delta_l$. For even orbital moments *l*

$$\cot\Delta_l = -\frac{\operatorname{Re}(a+b)}{\operatorname{Im}(a+b)} = \cot\eta_0, \tag{41}$$

and for odd *l*

$$\cot\Delta_l = -\frac{\operatorname{Re}(a-b)}{\operatorname{Im}(a-b)} = \cot\eta_1. \tag{42}$$

Here $\eta_0$ and $\eta_1$ are the molecular phases (22). The phases (41) and (42) are the exact solution of equation (39) but they are independent of $Y_{lm}(\mathbf{R})$. This dependence falls out while dividing equations (40). Thus, the necessary conditions to which the phase shifts $\Delta_l$ have to obey for equation (37) to be valid, namely: $\text{Im}\,\Delta_l = 0$ and $\Delta_l = \Delta_l[Y_{lm}(\mathbf{R})]$, cannot be met simultaneously, i.e. in the SSW approximation it is impossible to find the scattering phases that would provide the dependence of scattering amplitude on three main vector of the problem under consideration. Consequently, the assumption on the SSW asymptotic of the wave function cannot be considered as correct.

6. Numerical calculations

Bearing all of this in mind, the question arises: What electron energy should be for the results of calculation in the SSW approximation and in the partial wave method [20] to coincide? According to the general theory [20], this can be expected for low energy of a scattered particle. Compare the differential and total cross sections of elastic scattering within this range of energy.

To calculate the total cross section the even (41) and odd (42) phases with orbital moments $l \geq 2$ should be fallen out because for $k \to 0$ they do not obey the general law $\Delta_l(k) \to k^{2l+1}$. With taking this into account the total cross section in the SSW approximation is written as

$$\bar{\sigma}(k) = \frac{4\pi}{k^2} \sum_{l=0}^{\infty} (2l+1)\sin^2 \Delta_l = \frac{4\pi}{k^2}[\sin^2 \eta_0 + 3\sin^2 \eta_1] \qquad (43)$$

and it does not coincide with the exact cross section (31).

Note, that if we neglect the law on the behavior of phases for low energy and keep in the sum (43) all the phases (41) and (42), then we will get the following absurd result: the total cross section for slow electron scattering by two short-range potentials is infinite. Indeed, separating the even and odd $l$ terms in sum (43), we obtain

$$\bar{\sigma}(k) = \frac{4\pi}{k^2}\left[\sum_{l-even}^{\infty}(2l+1)\sin^2 \Delta_l + \sum_{l-odd}^{\infty}(2l+1)\sin^2 \Delta_l\right]. \qquad (44)$$

In this formula the squares of the sines can be factored out of the summations because they are constants for given $k$ and both sums are positive and infinite. It is evident that this absurd and non-coincidence of the general formulas obtained with asymptotic (32) with those based on the two-center asymptotic (9) is a consequence of the fact that for the non-spherical targets were used formulas (32), (36) and (43) correct for the spherical potentials only.

For low electron energy $k \to 0$ the second terms in the formulas (31) and (43) can be neglected since $\eta_0 \gg \eta_1$. In this limit the cross sections coincide. Thus, the SSW approximation leads to the correct result for the total cross section only when the electron wavelength significantly exceeds the target size, which is an agreement with the prediction of the general theory [20].

The differential cross section of elastic scattering is defined by the square of the amplitude module and has the form

$$\frac{d\sigma}{d\Omega} = |A(\mathbf{k}, \mathbf{k}')|^2 =$$
$$k^{-2}[\sin^2 \Delta_0 + 6\sin^2 \Delta_0 \sin^2 \Delta_1(1 + \cot \Delta_0 \cot \Delta_1)\cos\vartheta + 9\sin^2 \Delta_1 \cos^2 \vartheta] \qquad (45)$$

Here $\cos\vartheta = (\mathbf{k} \cdot \mathbf{k}')/k^2$ is the cosine of scattering angle. As in the case of the total cross section (43) we take into account here the first two phases only. The other phases should be neglected. Otherwise the amplitude of zero-angle scattering $A(\mathbf{k}, \mathbf{k})$ tends to infinity owing to the optical theorem [23]. For low electron energy $k \to 0$ the cross section (45) is isotropic and $d\sigma/d\Omega = k^{-2}\sin^2\Delta_0 = 4R^2(1+qR)^{-2}$. The exact amplitude of scattering tends to the same limit:

$$|F(\mathbf{k},\mathbf{k}',\mathbf{R})|^2_{k\to 0} = 4|a+b|^{-2}. \tag{46}$$

It is evident, however, that with the raise in electron energy the differential cross sections will be different at least because the cross section (45) is a function of scattering angle $\vartheta$ only, while $|F(\mathbf{k},\mathbf{k}',\mathbf{R})|^2$ is a function of angles between three vectors.

To illustrate these formulas we compare within the model of non-overlapping potential the cross sections obtained by the SSW approximation and by the method [20] for a $N_2$ molecule. The aim here is not calculations of the cross sections as such but the illustration of the differences in the results when the calculations by the both methods are performed with the same initial parameters of N atoms. Inside the atomic spheres (Regions $I_1$ and $I_2$ in Fig. 1*b*) according to [1] "the potential can be approximated by the sum of a central model potential… or could be derived from the molecular charge distribution". This potential defines the wave function and its logarithmic derivative on the atomic sphere surface and hence the phase shifts by this sphere. For low electron energy all the scattering characteristics are defined by the *s*-phase only. Choose this phase so that the total cross section of scattering by the molecule $N_2$ calculated according to Eq. (31) coincides with the experimental one. Then this phase is used to calculate the differential and total cross sections in the SSW approximation with the use Eq. (43) and Eq. (45). The range of electron energy within which the cross sections obtained by the both methods coincide will define the applicability range of the SSW approximation.

The results of such a calculation for $R$=1.094 Å [28] are given in Table 1. The first three columns are the electron speed $v$, electron wave vector $k$ and energy $\varepsilon$, respectively. The cross sections $\sigma_{\exp}$ were taken from the book [12] (Chapter 18). The fifth column is the calculation results $\bar{\sigma}$ according to formula (43). The parameter $q$ was chosen so that the cross section (31) coincides with the experimental one $\sigma_{\exp}$. The scattering phases $\delta_0$ for isolated atomic potential (the seventh column) correspond to these chosen values of $q = -k\cot\delta_0$. For comparison, the Hartree-Fock phases of scattering $\delta_0^{HF}$ by an isolated nitrogen atom calculated with the codes [29] are given in the last column. The difference between the phases $\delta_0$ and $\delta_0^{HF}$ is not great and it can be easily explained. It is evident that the phases of scattering by an isolated nitrogen atom and by the same atom in molecule $N_2$ have to be different and the closer they are to each other, the more correct the approximation of non-overlapping atomic potentials is. As it follows from Table 1, for energies $\varepsilon > 2$ eV the total cross sections (31) and (43) greatly differ.

The differential cross sections are much more sensitive to calculation details than the total ones. This can be seen in Figs. 2 and 3 where the cross sections calculated with the formulas (11) and (45) for electron energy $\varepsilon = 0.25$ eV and 1 eV are given. For these values of energy the total cross sections practically coincide. The scattering cross sections as a function of polar angle between the wave vectors $\mathbf{k}$ and $\mathbf{k}$' are presented in figures. The axis Z of the spherical coordinate system is supposed to coincide with the vector $\mathbf{k}$. The vectors $\mathbf{k}$' and $\mathbf{R}$ are in the plane XZ. The polar angle of the vector $\mathbf{R}$ is $\vartheta_R = 0°$, 30°, 60° and 90°. The shapes of the curves in the both figures are practically the same but their scale is different. If for the scattering angle $\vartheta = 0$ and $\vartheta_R = 90°$ the difference between the cross sections for $\varepsilon = 0.25$ eV is ~0.6 a.u., then for energy $\varepsilon = 1$eV it is greater than one order. It is seen that the electron angular distribution calculated with the formulas (11) and (45) are close only when the molecular axis coincides with the direction of the incident electron momentum. In the rest considered cases the polar angle dependencies strongly differ. For $\vartheta_R = 30°$ and 60° the molecular axis is in the right semi-plane ZX and the target is non-symmetrical relative to the direction of incident electron motion – the atom at the coordinate origin is a kind of a screen for the second atom. Therefore, the polar diagram in the right semi-plane ($0 \leq \vartheta \leq \pi$) greatly differs from that in the left semi-plane ZX ($\pi \leq \vartheta \leq 2\pi$). For $\vartheta_R = 90^o$ this asymmetry disappears since the both atoms are disposed symmetrically relative to the vector **k**. The calculations show that the differential cross sections (11) and (45) as a function of azimuth angle of the vector **k'** are also different.

7. Conclusions

It has been demonstrated that the assumption on the SSW asymptotic of the wave function of the molecular continuum cannot be considered as correct. The SSW assumption not only, in principle, excludes the phenomenon of particle diffraction by a multicenter target as manifestation of interference of secondary spherical waves formed in the process of scattering but also leads to incorrect numerical results, especially for the differential cross sections. It has been shown that the method of partial waves for non-spherical targets ought to be constructed according to the general theory [20] rather than by means of straightforward application of the usual S-matrix method developed for spherically symmetrical potentials.

The use of non-overlapping atomic potentials or muffin-tin-potentials for molecule description assumes that the wave functions inside atomic spheres (Regions $I_1$ and $I_2$) are known. Hence, one knows the sets of the scattering phases for each of the atomic potentials. In this case, as shown in [25], it is possible to obtain the elastic scattering amplitude in closed form and the calculation of this amplitude reduces to solving a system of non-homogeneous algebraic equations. Therefore, within the model of the non-overlapping atomic potentials there is no necessity to resort to the method of partial waves.

Acknowledgements
This work was supported by Uzbek Foundation Award Ф-2-1-12.

Table 1. Comparison of the total cross sections

| $v$, eV$^{1/2}$ | $k$, a.u. | $\varepsilon$, eV | $\sigma_{exp}$, a.u. | $\bar{\sigma}$, a.u. | $q$, a.u | $\delta_0$, rad | $\delta_0^{HF}$, rad |
|---|---|---|---|---|---|---|---|
| 0.5 | 0.1356 | 0.25 | 35.80 | 35.90 | 0.682 | 6.0869 | 6.0217 |
| 1.0 | 0.2712 | 1.0 | 37.01 | 38.67 | 0.610 | 5.8648 | 5.7660 |
| 1.5 | 0.4067 | 2.25 | 72.51 | 134.72 | 0.378 | 5.4612 | 5.5031 |
| 2.0 | 0.5423 | 4.0 | 46.47 | 75.81 | 0.358 | 5.2959 | 5.2893 |

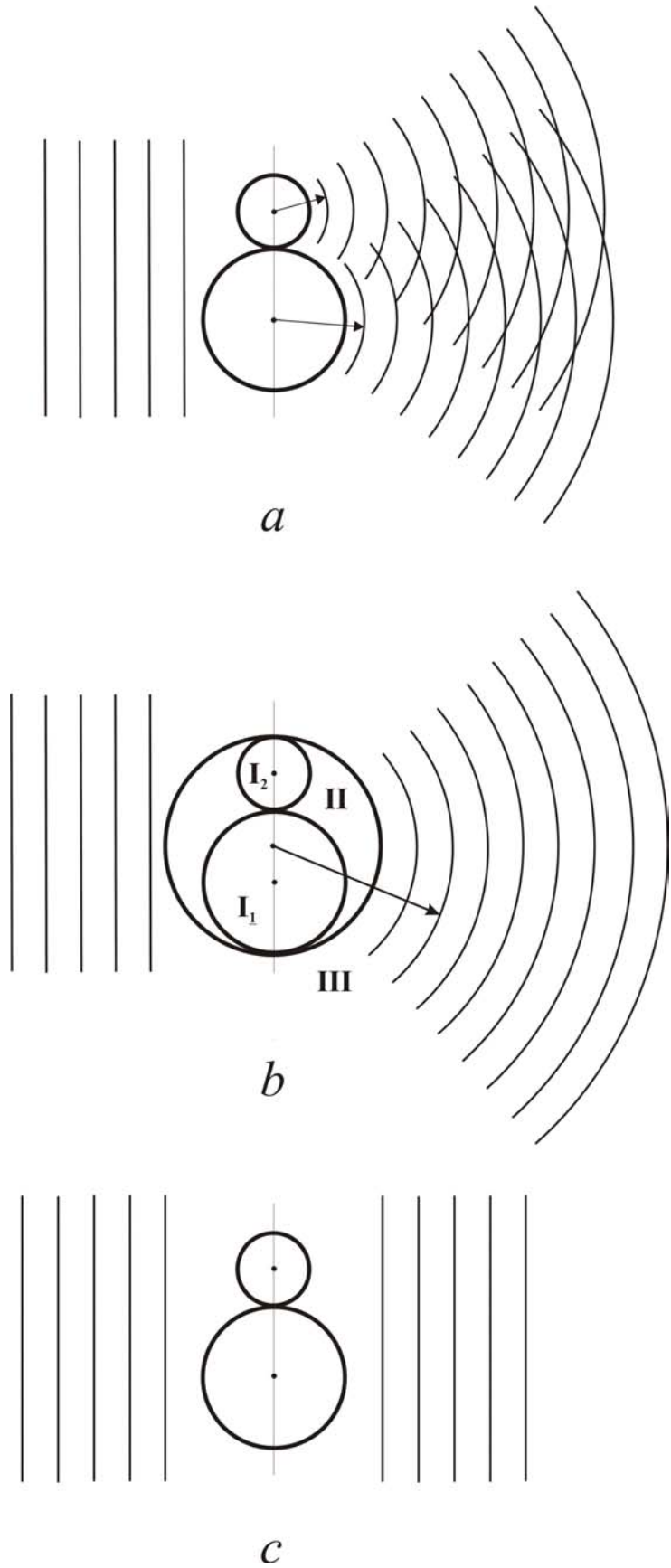

Fig. 1. *a* – the scattering picture according to the Huygens-Fresnel principle; *b* – Dill and Dehmer's scattering picture: *c* – scattering in the zeroth Born approximation [18].

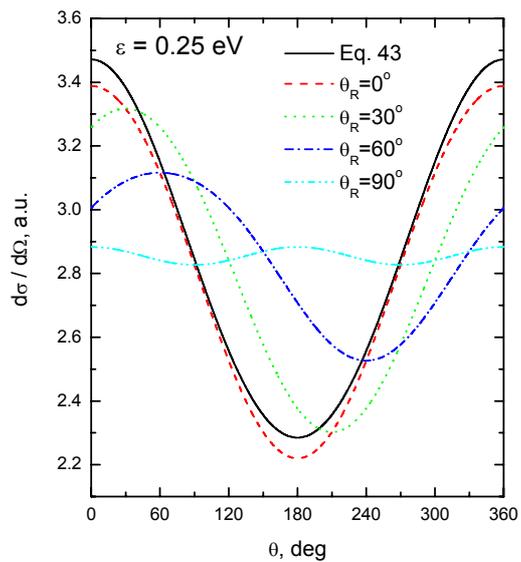

Fig. 2. The differential cross section of scattering for energy $\varepsilon = 0.25$ eV. The solid line is the SSW approximation (equation 43). The other lines are calculated with the amplitude (11) for different angles between the electron wave vector **k** and the molecular axis **R**.

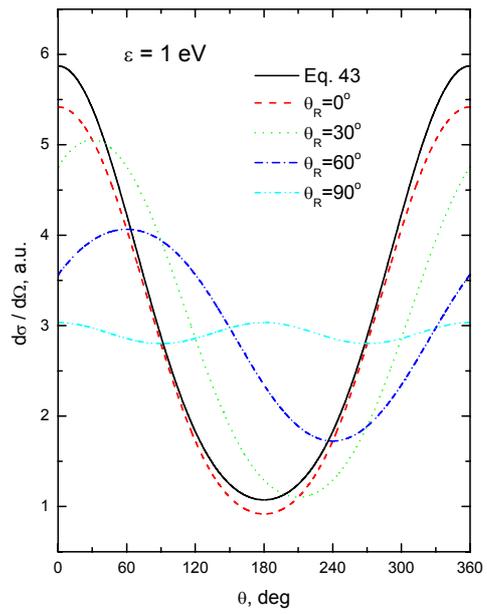

Fig. 3. The same as in Fig. 2 but for electron energy $\varepsilon = 1.0$ eV.